\documentclass[11pt]{article}
\usepackage[utf8]{inputenc}
\usepackage[T1]{fontenc}
\usepackage[english]{babel}
\usepackage{amsmath}
\usepackage{amssymb}
\usepackage{hep}
\usepackage{units}
\usepackage{dsfont}
\usepackage{mathrsfs}
\usepackage{feynmp}
\usepackage{graphicx}
\usepackage{float}
\usepackage{url}
\begin{document}

\begin{titlepage}
\hfill FTUV  10-0301

\vspace{1.5cm}
\begin{center}
{\bf\Large $\eta-\eta^\prime$- glueball mixing}\\[0.5cm]

\vskip 0.5cm

Simon Kiesewetter and Vicente Vento \\[0.5cm]

Departament de F\'{\i}sica Te\`orica and Institut de F\'{\i}sica Corpuscular,\\
Universitat de Val\`encia-CSIC, E-46100 Burjassot (Valencia), Spain.\\[1.0cm]

\end{center}

\begin{abstract}
We have revisited  glueball mixing with the pseudoscalar mesons in the MIT bag model scheme.
The calculation has been performed in the spherical cavity approximation to the bag using two
different fermion propagators, the cavity  and the free propagators. We obtain a very small probability
of mixing for the $\eta$ at the level of $0.04-0.1\%$ and a bigger for the $\eta'$ at the level of $4-12\%$. 
Our results differ from previous calculations in the same scheme
but seem to agree with the experimental analysis. We discuss the origin of our difference which
stems from the treatment of our time integrations.

\end{abstract}

\vskip 1.0cm
\noindent Pacs: 12.39.Mk,12.39.Ba,14.40.-n,14.65.Bt \\



\vfill

\noindent Email: name.surname@uv.es

\end{titlepage}

\section{Introduction}

Quantum Chromodynamics (QCD) is the theory of the hadronic interactions. It is an elegant
theory whose full non perturbative solution has escaped our knowledge since its formulation
more than 30 years ago\cite{FritzschGellMannLeutwyler}. The theory is asymptotically
free\cite{Gross:1973id,Politzer:1973fx} and confining\cite{Wilson:1974sk}. A particularly good
test of our understanding of the nonperturbative aspects of QCD is to study particles where
the gauge field plays a more important dynamical role than in the standard hadrons. For this
reason the glueball spectrum has attracted much attention ~\cite{Mathieu:2008me}. The interest
in this subject is related to the  significant progress in the understanding  of the properties
of such states within QCD, as well as, in the new possibilities for their identification in
modern experiments.

From the phenomenological point of view it has become clear by now that it is difficult to
single out which states of the hadronic spectrum are glueballs because we lack the necessary
knowledge to determine their decay properties. Moreover the strong expected mixing between
glueballs and quark states leads to a broadening of the possible glueball states which does
not simplify their isolation. The wishful sharp resonances which would confer the glueball
spectra the beauty and richness of the baryonic and mesonic spectra are lacking. This
confusing picture has led to a loss of theoretical and experimental interest in these hadronic
states. However, it is important to stress, that if they were to exist they would be a
beautiful and unique consequence of QCD. A comprehensive review on the experimental status
of glueballs has recently appeared \cite{Crede:2008vw}.

For the purposes of this paper we accept the existence of at least one pseudoscalar glueball state
although its existence has been a matter of debate since the Mark II experiment proposed
glueball candidates \cite{Scharre:1980zh}. Note that the pseudoscalar sector is a complex one.
On the one hand it accommodates the Goldstone nature of the pseudoscalar multiplet, on the other,
not totally unrelated, we encounter the singlet-octet mixing, which is traditionally associated with the
resolution of $U(1)$ anomaly. In constituent models the ideal mixing ($\theta_i= \tan^{-1}\sqrt{2}$) is
natural, however the $\eta$ and  $\eta^\prime$ mixing is non ideal.

Gluon self-couplings in QCD suggest the existence of glueballs, bound states of mainly gluons
\cite{FritzschMinkowski}. Investigating glueball physics requires an intimate knowledge of the
confining QCD vacuum and it is well known that such properties cannot be obtained using standard
perturbative techniques. To handle the non-perturbative regime of QCD, one can resort to numerical
methods, known as lattice QCD. Lattice QCD needs as input the quark masses and an overall scale,
conventionally given by $\Lambda_{QCD}$. Then any Green function can be evaluated by taking average
of suitable combinations of lattice fields in vacuum samples. This allows masses  and matrix elements,
particularly those of weak or electromagnetic currents, to be studied. However lattice QCD faces both
computational and fundamental problems in the description of glueballs \cite{Mathieu:2008me}.
A complementary way to describe glueballs, namely the MIT bag model, implements in a dynamical way the
phenomenological properties of the confining QCD vacuum and the interaction among the gluons.
Historically the investigation of the glueball properties started precisely in this model
\cite{Jaffe:1975fd}.  Jaffe and Johnson found many glueball states with different quantum numbers lying
in the mass interval 1000-2000 MeV. They emphasized that one should expect rather small widths for such
states because their decays in conventional hadrons violate the Okubo-Zweig-Iizuka  (OZI) rule \cite{ozi}.

The aim of the present investigation is  the study of the mixing between a possible pseudoscalar glueball
state and the $\eta$ or $\eta'$-mesons. The calculation has been performed in the MIT bag model, a description
which imposes by fiat some of the properties of QCD. In this model a hadron is basically a bubble of
perturbative vacuum in the midst of a non-perturbative vacuum. Inside the bubble we insert the constituents,
which are described by cavity modes, and the surface of the bubble screens color from flowing into the
non-perturbative world. The calculation has been performed in the so-called spherical cavity approximation,
where several improvements have been incorporated, like center-of-mass corrections and the recoil correction.
In this setup, the cavity is fixed to be a sphere and its radius is allowed to vary dynamically. Within this
scheme we have performed two calculations of the mixing Hamiltonian. One, in which we have used the cavity
propagator for the quarks . This cavity propagator is made up of a sum over all possible cavity states. Thus
it incorporates, in principle, the confining property of the bag model. Another, in which we have used the free 
propagator which is made up as a sum over free modes. As it turns out, the results of both calculations are
almost the same, so the dominating property, at least for the problem investigated here, is asymptotic freedom.

Our investigation is presented as follows. In \ref{mixings} we show the necessary tools to carry out the calculation.
Starting from the QCD Lagrangian we use a formalism that allows one to calculate the mixing energies perturbatively
by means of the appropriate Feynman diagrams. We have to introduce for this purpose the bare glueball and meson states. 
In sec. \ref{propagators}, we discuss the quantization, which is important since the problem at hands is a multi-particle one. 
After discussing  the role of the propagator in sec. \ref{propagators}, and addressing and resolving an important physical problem 
that arises in bag model calculations in sec. \ref{recoil}, we present and comment on the results in sec. \ref{results} and give some 
conclusions in sec. \ref{conclusion}.  The actual calculations have been relegated to the appendix to ease the reading of the main text.

\section{Calculation of the $\eta$-$\eta'$-glueball mixing in the bag model}
\label{mixings}

We next calculate the mixing energy, which corresponds to off-diagonal Hamiltonian terms in Fock space.
In  subsection \ref{formalism}, we introduce a formalism which allows the calculation of the mixing energies
in a perturbative manner. In subsection \ref{states} we present the bare glueball and meson states in the
bag model. Thereafter we discuss important aspects of the calculation

\subsection{Formalism}\label{formalism}

QCD is a non-abelian Yang-Mills theory with a $\mathcal{S}\mathcal{U}(3)$ gauge symmetry regarding color charge. The Lagrangian is
\begin{align}
\mathcal{L}_{QCD}=\overline{\psi}\left(i\slashed{D}-m\right)\psi-\frac{1}{4}F_{\mu\nu}^aF^{\mu\nu}_a
\end{align}
where
\begin{eqnarray}
D_\mu &= &\partial_\mu-igA_\mu^a t^a\\
F_{\mu\nu}^a & = &\partial_\mu A_\mu^a-\partial_\nu A_\mu^a + g f^{abc}A_\mu^b A_\nu^c
\end{eqnarray}
$t^a=\frac{1}{2}\lambda^a$ where $\lambda^a$ are the Gell-Mann matrices and $f^{abc}$ are the structure constants of the $SU(3)$ algebra. Some rearrangement yields
\begin{eqnarray}
\mathcal{L}_{QCD} & = &\underbrace{\overline{\psi}\left(i\slashed{\partial}-m\right)\psi}_{\mathcal{L}_{0_A}}-\underbrace{\frac{1}{4}\left(\partial_\mu A_\nu^a-\partial_\nu A_\mu^a\right)\left(\partial^\mu A_a^\nu-\partial^\nu A_a^\mu\right)}_{\mathcal{L}_{0_B}} \nonumber\\
&&+\underbrace{g\overline{\psi}\gamma^\mu A_\mu^a t^a \psi}_{\mathcal{L}_1} - \underbrace{g\left(\partial_\mu A_\nu^a\right)f^{abc}A_b^\mu A_c^\nu}_{\mathcal{L}_2} - \underbrace{\frac{1}{4}g^2\left(f^{abc}A_\mu^bA_\nu^c\right)\left(f^{abc}A_b^\mu A_c^\nu\right)}_{\mathcal{L}_3}
\end{eqnarray}
where we can identify the free Lagrangian of QED ($\mathcal{L}_0$) with color indices, a quark-quark-gluon vertex ($\mathcal{L}_1$), a 3-gluon vertex ($\mathcal{L}_2$) and a 4-gluon vertex ($\mathcal{L}_3$).

\begin{figure}
\centering
\includegraphics[angle=0,scale=0.4]{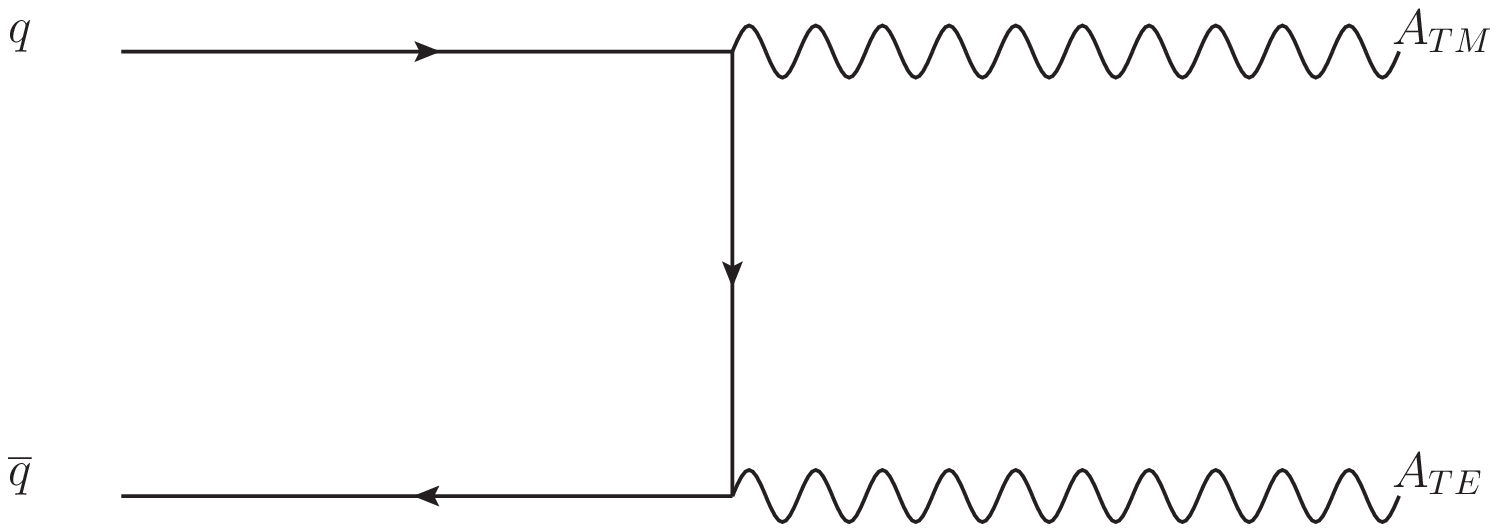}
\includegraphics[angle=0,scale=0.4]{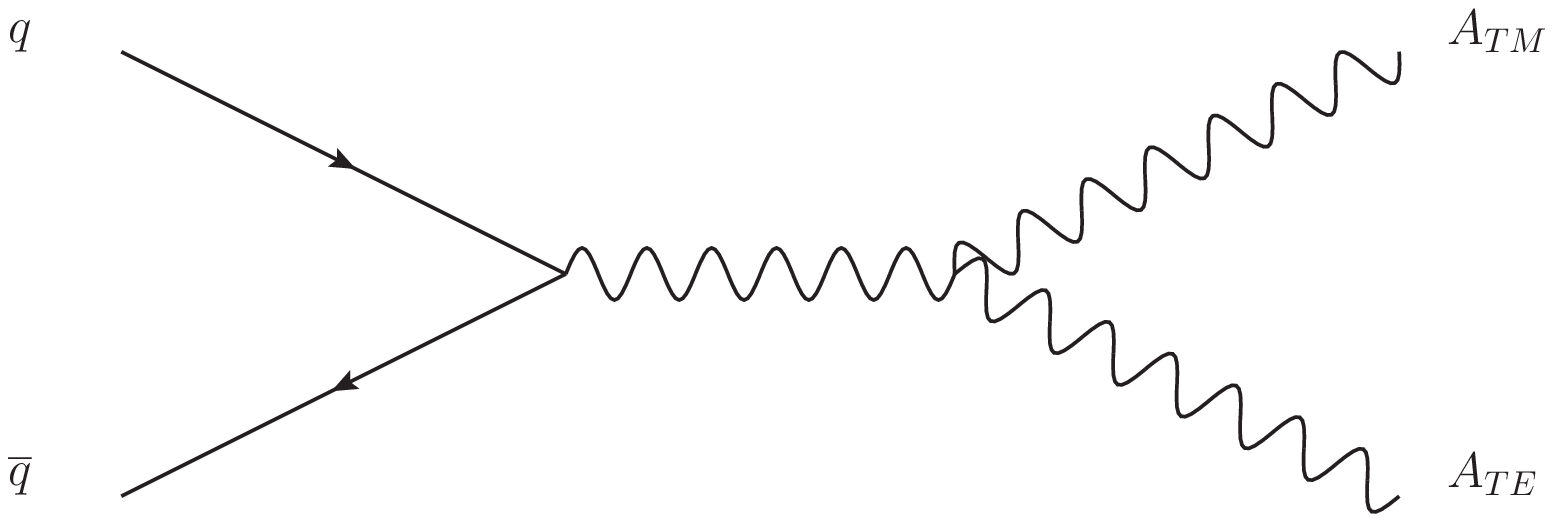}
\caption{Fermion exchange contribution to the meson-glueball mixing (left) and
Gluon exchange to lowest order gives no contribution(right).}
\label{fgexchange}
\end{figure}

We use  a perturbative approach inside the cavity following the scheme developed by Maxwell and Vento\cite{Maxwell:1981kg}. The glueball and meson states represent solutions of the free Lagrangian $\mathcal{L}_{0_A}$ and $\mathcal{L}_{0_B}$, respectively. The equations of motion arising from the Lagrangian of QCD are
\begin{align}
\left(i\slashed{\partial}-m\right)\psi=g\gamma^\mu A_\mu^a t^a \psi \label{eq-thbg-formal-eqmoquark}
\end{align}
and
\begin{align}
\partial_\mu\left(\partial^\mu A_a^\nu-\partial^\nu A_a^\mu\right)=-gf^{abc}\partial_\mu\left(A_b^\mu A_c^\nu\right)-gf^{abc}A_\mu^b F^{\mu\nu}_c - g\overline{\psi}\gamma^\nu t_a \psi\label{eq-thbg-formal-eqmogluon}
\end{align}
Eqs.\eqref{eq-thbg-formal-eqmoquark} and \eqref{eq-thbg-formal-eqmogluon} can be understood as an inhomogenous Dirac equation and Maxwell equation, respectively. Thus, they can be solved exactly using the Feynman propagator for the Dirac field and the Maxwell field, respectively in the following way
\begin{align}
\psi(x)=g \int d^4x' S_F(x,x') \gamma^\mu A_\mu^a(x') t^a \psi(x')
\end{align}
and
\begin{align}
A_a^\mu(x) = g \int d^4x' D_F(x,x') \left[-f^{abc}\partial_\mu\left(A_b^\mu(x') A_c^\nu(x')\right)-f^{abc}A_\mu^b(x') F^{\mu\nu}_c(x')-\overline{\psi}\gamma^\nu t_a \psi\right]
\end{align}
One can now expand $\psi(x)$ and $A^\mu(x)$ in a power series of $g$ and obtains for the first order term
\begin{align}
\psi^{(1)}(x)=\psi^{(0)}+g \int d^4x' S_F(x,x') \gamma^\mu A^{(0)a}_\mu(x') t^a \psi^{(0)}(x')
\end{align}
\begin{align}
A^{(1)\mu}_a(x) = A^{(0)}+g \int d^4x' D_F(x,x') \left[-f^{abc}\partial_\mu\left(A^{(0)\mu}_b(x') A^{(0)\nu}_c(x')\right)\nonumber\right.\\
\left. \vphantom{\left(A^{(0)\mu}_b(x') A^{(0)\nu}_c(x')\right)} +f^{abc}A^{(0)b}_\mu(x') F^{(0)\mu\nu}_c(x')\right]
\end{align}
One can now use the expressions $\psi^{(1)}$ and $A^{(1)}$ or higher orders of $\psi$ and $A$ to calculate the expectation value $\left<\hat{\Gamma}\right>$ of some observable $\hat{\Gamma}$ to various perturbative orders of $g$. In here we are interested in the expectation value of the quark-gluon interaction Hamiltonian $\hat{H}_I=g\overline{\psi}\gamma^\mu A_\mu^a t^a \psi$. Inserting $\psi^{(1)}$ yields
\begin{eqnarray}
\left<\overline{H_I}\right> & = &g\int d^3x \overline{\psi}^{(1)}(x)\gamma^\mu A^{(0)}(x)_\mu^at^a\psi^{(0)}(x) \nonumber\\
& & +g\int d^3x \overline{\psi}^{(0)}(x)\gamma^\mu A^{(0)}(x)_\mu^at^a\psi^{(1)}(x) \\
& = & 2g^2\int d^3x \overline{\psi}(x)\gamma^\mu A(x)_\mu^at^a\int d^4x' S_F(x,x') \gamma^\nu A(x')_\nu^b t^b \psi(x')
\label{eq-thbg-formal-matel}
\end{eqnarray}
with zero order wavefunctions in eq.\eqref{eq-thbg-formal-matel}. This expression corresponds to an exchange of a virtual fermion as shown in fig. \ref{fgexchange}. Inserting different orders of $\psi$ and $A$, one obtains expressions corresponding to different Feynman diagrams. Every order in $\psi$ brings a fermion propagator, while every order in $A$ brings a gluon propagator. For our calculation, we will restrict ourselves to the one-fermion exchange, since this is the leading-order diagram of the meson-gluon interactions. The lowest-order gluon exchange diagram, shown in fig. \ref{fgexchange}, does not contribute, because the gluon is a spin-1 particle and therefore does not couple to the spin-0 pseudoscalar meson or glueball.

\subsection{Glueball and meson states}\label{states}
In order to construct the glueball states, we have to describe the gluon cavity modes. The 4-vector potential is given by
\begin{align}
A^\mu = \left(-A^0\equiv\phi,\vec{A}\right)
\end{align}
Since we are working in the static cavity approximation, we will be using the Coulomb gauge
\begin{align}
\vec{\nabla}\cdot\vec{A} = 0
\end{align}
In this gauge, the scalar potential is given by Poisson's equation
\begin{align}
-\Delta \phi = \frac{\rho}{\epsilon_0}
\end{align}
and thus vanishes since there are no free charges in the model considered here.\\

The solutions can be classified into two different classes, which are called transverse electric (TE) and transverse magnetic (TM).

Furthermore, the solutions are classified by the quantum numbers
\begin{align}
l={}&1,2,... \text{ orbital excitation}\\
m={}&-1,+1 \text{ magnetic quantum number}
\end{align}
The boundary conditions generate the constraints
\begin{align}
\left.\frac{d}{dr}\left(rj_l\left(kr\right)\right)\right|_{r=R}=0 \label{eq-states-tebc1}
\end{align}
for the TE solution and
\begin{align}
j_l(kR)=0 \label{eq-states-tmbc1}
\end{align}
for the TM solutions with $j_l$ being a spherical bessel function of order $l$. The transcendental equations \eqref{eq-states-tebc1} and \eqref{eq-states-tmbc1} have an infinite number of solutions $kR=x_n$, labeled by the radial quantum number $n=0,1,...$.  The lowest modes of interest here are l=1, n=0, $x_{TE}= 2.74$ and $x_{TM}= 4.49$ \cite{Maxwell:1981kg}.

The parities of the modes are $\pi=(-1)^{l+1}$ for the TE modes and $\pi=(-1)^l$ for the TM modes. The non-linear boundary condition requires $l=0$. This, however, is incompatible with the helicity of the gluon. Thus, we must assume that the best value for $l$ is the lowest possible, that is $l=1$. The lowest-lying pseudoscalar glueball with parity $J^{\pi C}=0^{-+}$, which is the objective of our investigation, contains the lowest-lying TE mode and TM mode gluon.

Let $a_{lkmn}^\dagger$ denote the particle creation  operator associated with the gluon cavity-state denoted by the quantum numbers $l,\kappa,m,n$, where $\kappa\in\left\{TE,TM\right\}$ denotes the polarisation. The lowest-lying state with $\kappa=TE$ is given by $l=1,n=0$ and the lowest-lying state with $\kappa=TM$ is given by $l=1,n=0$. Thus, the glueball state can be constructed by
\begin{align}
\ket{G}=\frac{1}{\sqrt{2}}\left(\hat{a}_{TE\uparrow}^\dagger \hat{a}_{TM\downarrow}^\dagger-\hat{a}_{TE\downarrow}^\dagger \hat{a}_{TM\uparrow}^\dagger\right)\ket{0} \label{eq-states-glueball-operators}
\end{align}
where $l=1,n=0$ everywhere. We impose the restriction that the state be a color singlet.\\

The meson states are constructed from the cavity fermion modes. To find them, we have to study the radial solutions of the free Dirac equation. They are characterized by their total angular momentum $j=\nicefrac{1}{2},\nicefrac{3}{2},...$, a magnetic quantum number $m=-\nicefrac{1}{2},\nicefrac{1}{2}$ and another quantum number $\lambda=-1,1$, called Dirac's quantum number. The wavefunctions are\cite{Messiah62}
\begin{align}
u\left(x\right)&=-N\begin{pmatrix}i\lambda j_l\left(pr\right) \\ \Omega j_{l'}\left(pr\right)\left(\vec{\sigma}\cdot\hat{\vec{x}}\right)\end{pmatrix}\mathscr{Y}_{lj}^{m}(\hat{x})e^{-i\omega t} \label{eq-states-quark-u} \\
v\left(x\right)&=N\begin{pmatrix}i\Omega j_{l'}\left(pr\right)\left(\vec{\sigma}\cdot\hat{\vec{x}}\right)\\\lambda j_l\left(pr\right)\end{pmatrix}\mathscr{Y}_{lj}^{m}(\hat{x})e^{i\omega t} \label{eq-states-quark-v}
\end{align}
where $l\equiv J+\frac{1}{2}\lambda$ and $l'\equiv J-\frac{1}{2}\lambda$, $\Omega\equiv \frac{p}{\omega+m}$. In the case of a massive field, $p$ and $\omega$ are related by $\omega=\sqrt{p^2+m^2}$, otherwise $p\equiv\omega$.
Here $u$ denotes the particle solution and $v$ the antiparticle solution.
The object $\mathscr{Y}_{lj}^{m}$ is a 2-spinor of total angular momentum $j$, projection $m$ and orbital angular momentum $l$, called spinor spherical harmonics, defined by
\begin{align}
\mathscr{Y}_{lJ}^{M}(\hat{x})\equiv \sum_{m\mu}\left<\begin{array}{cccc|cc}lm\frac{1}{2}\mu|JM\end{array}\right>Y_{lm}\left(\hat{x}\right)\chi_{\mu}
\end{align}
Inserting eq.\eqref{eq-states-quark-u} and eq.\eqref{eq-states-quark-v} into the boundary condition yields the constraint
\begin{align}
j_{l'}\left(pR\right)=-\frac{\lambda}{\Omega}j_l(pR)
\end{align}
This transcendental equation has an infinite set of solutions for each combination of $j, \lambda$. The non-linear boundary condition requires $j=\frac{1}{2}$ for the quark states.
The quark states are normalized by the requirement
\begin{align}
\int_V d^3x u^\dagger(x) u(x) = 1.
\end{align}
This yields for the lowest-lying state ($j=\frac{1}{2},n=0,\lambda=-1$) a normalization constant of
\begin{align}
\label{norm}
N^2=\frac{1}{R^3j_0(\tilde{x})^2} \frac{\tilde{x}\left(\tilde{\nu}-mR\right)}{2\tilde{x}\left(\tilde{\nu}-1\right)+mR}\frac{1}{4\pi}.
\end{align}
where $\tilde{x}\equiv\omega R$ and $\tilde{\nu}\equiv p R$.
Note that this factor differs from the one used in~\cite{DeGrand:1975cf} and thereafter by many authors. We have checked numerically that this is the correct formula.

The $\eta$-mesons are pseudoscalar mesons, thus have parity $J^{PC}=0^{-+}$. Let $\hat{b}_{j\lambda mn}^\dagger$ and $\hat{d}_{j\lambda mn}^\dagger$ denote the particle creation operator and anti-particle creation operator, respectively associated with the fermion cavity-state denoted by the quantum numbers $j,\lambda,m,n$. We are interested in the lowest-lying meson state. The lowest energy modes are associated to the quantum numbers $j=\frac{1}{2},n=0,\lambda=-1$. Because of intrinsic parity between particle and antiparticle states, $P=-1$ can be obtained using the lowest-lying particle state and the lowest-lying antiparticle state. Thus, the meson state can be constructed by
\begin{align}
\ket{\sigma}=\frac{1}{\sqrt{2}}\left(\hat{b}_{\uparrow}^\dagger \hat{d}_{\uparrow}^\dagger-\hat{b}_{\downarrow}^\dagger \hat{d}_{\downarrow}^\dagger\right)\ket{0} \label{eq-states-meson-operators}
\end{align}
where $j=\frac{1}{2},n=0,\lambda=-1$ everywhere. We have used the notation $\uparrow,\downarrow$ for $m=\frac{1}{2},-\frac{1}{2}$\\
Furthermore, we impose a color singlet state and a flavor composition of
\begin{align}
\eta =  \frac{1}{\sqrt{6}}\left(u\overline{u}+d\overline{d}-2s\overline{s}\right)  \\
\eta^\prime = \frac{1}{\sqrt{3}}\left(u\overline{u}+d\overline{d}+s\overline{s}\right).
\end{align}

The above definition of $\eta$ and $\eta'$ is actually an approximation, since we know that due to the chiral anomaly the physical $\eta$ and $\eta'$ states are mixtures with a small mixing angle of $\eta_1$ and $\eta_8$. We  neglect this fact throughout this work and identify $\eta$ with $\eta_8$ and $\eta'$ with $\eta_1$, as done above.

\subsection{Propagators}\label{propagators}
The off-diagonal term in the energy expectation value is given by
\begin{align}
\bra{G}N\left\{2g^2\int d^3x \overline{\psi}(x)\gamma^\mu A(x)_\mu^at^a\int d^4x' S_F(x,x') \gamma^\nu A(x')_\nu^b t^b \psi(x')\right\}\ket{\sigma} \label{eq-thbg-quantiz-matel}
\end{align}
where $N$ denotes the normal ordering operator.
This expression can be evaluated using eq.\eqref{eq-states-meson-operators} and eq.\eqref{eq-states-glueball-operators} when one quantizes eq.\eqref{eq-thbg-formal-matel} in the manner
\begin{align}
\psi(x)\rightarrow\hat{\psi}(x)={}&\sum_\alpha\left[u_\alpha(x)\hat{b}_\alpha+v_\alpha(x)\hat{d}_\alpha^\dagger\right]\\
\overline{\psi}(x)\rightarrow\hat{\overline{\psi}}(x)={}&\sum_\alpha\left[\overline{v}_\alpha(x)\hat{d}_\alpha+\overline{u}_\alpha(x)\hat{b}_\alpha^\dagger\right]\\
A(x)\rightarrow\hat{A}(x)={}&\sum_\alpha\left[A_\alpha(x)\hat{a}_\alpha+A_\alpha^*(x)\hat{a}_\alpha^\dagger\right]
\end{align}
with the commutation relations
\begin{align}
\text{ Fermion:   }  \left\{\hat{b}_\alpha,\hat{b}_\beta^\dagger\right\}={}&\delta_{\alpha\beta}\\
\text{ Fermion:   } \left\{\hat{d}_\alpha,\hat{d}_\beta^\dagger\right\}={}&\delta_{\alpha\beta} \\
\text{ Gluon:      } \left[\hat{a}_\alpha,\hat{a}_\beta^\dagger\right]={}&\delta_{\alpha\beta}
\end{align}
Evaluating eq.\eqref{eq-thbg-quantiz-matel} with a normal-ordered operator and shifting $\hat{b}^\dagger$, $\hat{d}^\dagger$ and $\hat{a}^\dagger$ to the left and $\hat{b}$, $\hat{d}$ and $\hat{a}$ to the right yields
\begin{align}
\text{Fermion:    } \delta_{\alpha\uparrow}\delta_{\beta\uparrow}-\delta_{\alpha\downarrow}\delta_{\beta\downarrow}
\end{align}
\begin{align}
\text{Gluon:     } \delta_{\iota TE\uparrow}\delta_{\kappa TM\downarrow}-\delta_{\iota TE\downarrow}\delta_{\kappa TM\uparrow}+\delta_{\iota TM\downarrow}\delta_{\kappa TE\uparrow} - \delta_{\iota TM\uparrow}\delta_{\kappa TE \downarrow} ,
\end{align}
where $\alpha$ refers to the sum index in $\overline{\psi}$ with {\it particle} states, $\beta$ to the sum index in $\psi$ with {\it antiparticle} states, $\iota$ to the sum index in the first gluon wavefunction $A$ and $\kappa$ to the sum index in the second gluon wavefunction $A$.

There are two possible choices for the propagator $S_F(x,x')$ in eq.\eqref{eq-thbg-formal-matel}, namely the confined propagator and the free propagator. The confined propagator is built of a complete set of confined states in the bag model. The free propagator is the well-known Feynman propagator $S_F(x,x')=\int \frac{d^4p}{(2\pi)^4}e^{-ip\cdot(x-x')}\frac{1}{\slashed{p}-m+i\epsilon}$ with the Feynman prescription of closing the contour. Both propagators act as Green's functions with respect to the Dirac operator, thus formally both propagators can be used. However, since they are of a very different shape, one should expect them to generate different results. Physically, it is not clear which propagator is preferable. On the one hand, virtual fermion are not constrained to the bag as the bound states are. On the other hand, the bound states are not momentum eigenstates, so the problem does not resemble a scattering process and there is no 4-momentum conservation at the vertices. These problematic will be elaborated in more detailed in sec. \ref{recoil} and with the recoil correction, a possible resolution is presented. In a way, the confined propagator may overemphasize the aspect of confinement, while the free propagator may overemphasize the aspect of asymptotic freedom in the bag model. Out of curiosity, we will carry out the calculation using both propagators.

\begin{enumerate}
\item  Confined propagator:
Since the radial solutions of the Dirac equation eq.\eqref{eq-states-quark-u} and eq.\eqref{eq-states-quark-v} form a complete set, we will follow an approach by Maxwell and Vento\cite{Maxwell:1981kg} and use them to construct the confined propagator
\begin{align}
-i S_F(x,x')={}&\sum_\alpha\left[u_\alpha(x)\overline{u}_\alpha(x')e^{-i\omega_\alpha(t-t')}\theta(t-t')\right.\nonumber\\
  &\left.-v_\alpha(x)\overline{v}_\alpha(x')e^{i\omega_\alpha(t-t')}\theta(t'-t)\right] \label{eq-calc-conf-propagator}
\end{align}
with $\alpha=(n,\lambda,j,m)$ denoting a multiindex. Note that since the solutions represent virtual particles rather than real quark states, they are not subject to the non-linear boundary condition and thus, values other than $j=\frac{1}{2}$ are possible.

\item Free propagator:
The solutions generating the free propagator are not subject to the boundary conditions. Dropping the boundary conditions yields
\begin{align}
\sum_{n,\lambda,J,M}\rightarrow\displaystyle\int_k dk \sum_{\lambda,J,M}
\end{align}
Imposing the normalization condition
\begin{align}
\int d^3x u_k^\dagger(x)u_{k'}(x)=\int d^3x v_k(x)v_{k'}(x) = \delta(k-k')
\end{align}
yields
\begin{align}
u_k(x)&= \sqrt{\frac{2}{\pi}}\frac{k}{\sqrt{1+\Omega^2}}\begin{pmatrix}i\lambda j_l\left(kr\right) \\ \Omega j_{l'}\left(kr\right)\left(\vec{\sigma}\cdot\hat{\vec{x}}\right)\end{pmatrix}\mathscr{Y}_{lJ}^{M}(\hat{x})e^{-i\omega t}\\
v_k(x)&= \sqrt{\frac{2}{\pi}}\frac{k}{\sqrt{1+\Omega^2}}\begin{pmatrix}i\Omega j_{l'}\left(pr\right)\left(\vec{\sigma}\cdot\hat{\vec{x}}\right)\\\lambda j_l\left(pr\right)\end{pmatrix}\mathscr{Y}_{lJ}^{M}(\hat{x})e^{i\omega t}
\end{align}
where we have used
\begin{align}
\int dr r^2 j_l(kr) j_l(k'r) = \frac{\pi}{2k^2}\delta(k-k')
\end{align}
This way, the calculation for the free propagator is analogous to the one shown in the appendix, with $\displaystyle\sum_n\rightarrow\int_k dk$ and a different normalization of $\psi_\alpha$. This corresponds directly to Rayleigh's expansion of plane waves.

\end{enumerate}

\subsection{Time integration and recoil correction}\label{recoil}
\begin{figure}
\centering
\includegraphics[angle=0,scale=0.6]{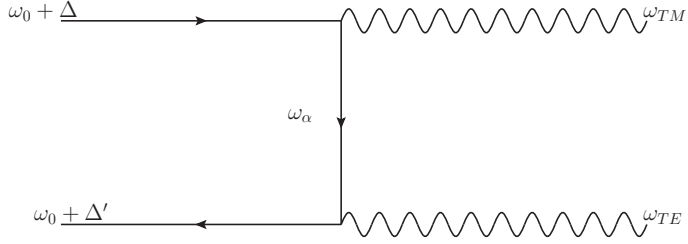}
\caption{Recoil correction; Energy $\Delta$ and $\Delta'$ has to be added to account for center-of-mass motion that necessarily arises through emission and absorption of the virtual particle}
\label{fig-recoil-corr}
\end{figure}
As mentioned above, energy conservation on the vertices is not possible in the process described here since $\omega_0\neq\omega_{TE}$ and $\omega_0\neq\omega_{TM}$. Thus, the integrations of eq.\eqref{eq-thbg-formal-matel} might be formally carried out (although for specific quark masses, there arise unphysical divergences), but the physical interpretation remains unclear. It is also unclear, why the $dt'$-integration (corresponding to the lower vertex) has to be carried out but the $dt$-intergration doesn't, so that there will be an energy denominator only related to the energies at the lower vertex. Apparently, there must be a (physical) flaw somewhere. In fact, the flaw lies within the bag. When virtual particles that violate energy conservation are being created, a center-of-mass motion must be expected that ultimately alters the energies of the states. This process will be taken into account here and we will call this the recoil correction. Through the emission and absorbtion of the virtual fermion the states will obtain an additional center-of-mass motion energy $\Delta$ and $\Delta'$, respectively as seen in fig. \ref{fig-recoil-corr}, so that we are in the center-of-energy frame. This gives
\begin{align}
\omega_0+\Delta+\omega_{TM}&=0\\
-\omega_0-\Delta'+\omega_{TE}&=0
\end{align}
Only in the center-of-energy frame can the virtual fermion exchange be understood in a physically plausible way. Making this approximation, my work differs from other works in the bag model, where this issue has not been addressed.

Writing down the time dependence, which we have left out before and applying the recoil correction yields
\begin{eqnarray}
&&\overline{u}_0(x)e^{i(\omega_0+\Delta)t}\slashed{A}_{TM}(x)e^{-i\omega_{TM}t}\int dt'\left[u_\alpha(x)\overline{u}_\alpha(x')e^{-i\omega_\alpha(t-t')}\theta(t-t')\right.\nonumber\\
&&\left.-v_\alpha(x)\overline{v}_\alpha(x')e^{i\omega_\alpha(t-t')}\theta(t'-t)\right]\slashed{A}_{TE}(x')e^{-i\omega_{TE}t'}v_0(x')e^{i(\omega_0+\Delta')t'}
\end{eqnarray}
where we have left out the spatial integrations for convenience. Performing the $dt'$ integration yields
\begin{eqnarray}
&&\int dt' e^{i\omega_\alpha t'}\theta(t-t')=\int_{-\infty}^t dt' e^{i\omega_\alpha t'}=\frac{1}{i\omega_\alpha}\left[e^{i\omega_\alpha t'}\right]_{-\infty}^t=\frac{e^{i\omega_\alpha t}}{i\omega_\alpha}\\
&&\int dt' e^{-i\omega_\alpha t'}\theta(t'-t)=\int^{\infty}_t dt' e^{-i\omega_\alpha t'}=-\frac{1}{i\omega_\alpha}\left[e^{-i\omega_\alpha t'}\right]^{\infty}_t=\frac{e^{-i\omega_\alpha t}}{i\omega_\alpha}
\end{eqnarray}
where we have shifted $\omega_\alpha\rightarrow\omega_\alpha-i\epsilon$ implicitly. This gives an overall denominator of $\frac{1}{i\omega_\alpha}$ for each mode which is being propagated

\section{Results}\label{results}

After carrying out a detailed calculation, which can be found in the appendix, one obtains for the mixing energy fig. \ref{final} as a function of the quark mass times the bag radius. The shown energy is per quark-pair, i.e. $q\bar{q}$. To calculate the corresponding mixing energy one has to take into account the wave function of the meson states.
As mentioned already in sec. \ref{states}, we are neglecting the $\eta-\eta'$-mixing and  identify $\eta$ as $\eta_8$ and $\eta'$ as $\eta_1$. Please note that both calculations, the one using the confined propagator, and the one using the free propagator, give very similar results. From now on we will only use the confined propagator results.

The values of the glueball mass change dramatically in the literature from one calculation to another. Lattice QCD in the quenched approximation leads to a value around $m_G =2500$ MeV \cite{Morningstar:1999rf,Meyer:2004jc,Chen:2005mg}. Unquenched calculations should produce a lower value as happens for the scalar glueball \cite{Sexton:1995kd}. This has been shown to be the case in an effective theory calculation of glueball mixing which reproduces a large amount of data \cite{Mathieu:2009sg},  where the lower pseudoscalar mass value is set at 2000 MeV. Other effective theory calculations which fit parameters to data  lead to values down to 1400 MeV \cite{Cheng:2008ss}.

\begin{figure}[h]
\begin{center}
\includegraphics[angle=-90,scale=0.4]{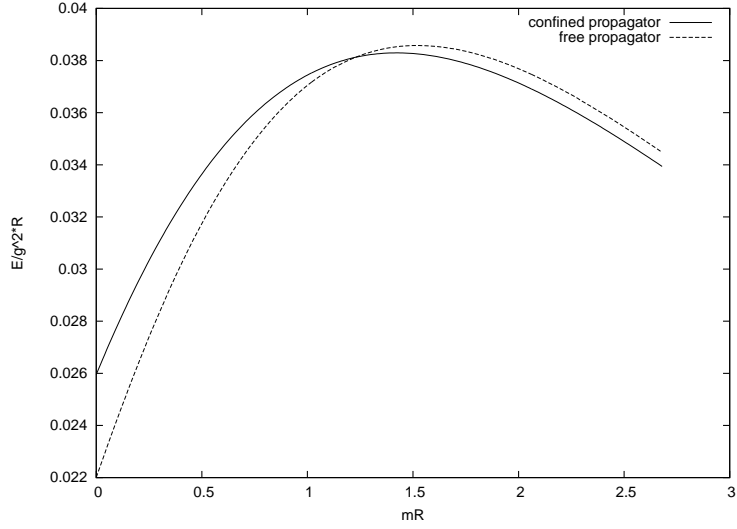}
\end{center}
\caption{Glueball-meson mixing energy per quark-antiquark pair as a function of $mR$ for the free and confined propagator}
\label{final}
\end{figure}

Kuti\cite{Kuti:1998rh} suggested that a reliable glueball spectrum, which is in reasonable agreement with lattice calculations, can be obtained for $B^{\nicefrac{1}{4}}=280$ MeV and $R\approx 0.5$ fm $=2.5$ GeV$^{-1}$. He gives a coupling constant $\alpha_S=0.5$ to obtains a $0^{-+}$ glueball mass of about $2500$ MeV in sharp contrast with the old calculation of Jaffe and Johnson \cite{Jaffe:1975fd}, who chose parameters closely related to the baryon spectrum. 
In order to perform our mixing calculation we shall take for the mesons the experimental values, $\eta(550)$ and $\eta'(960)$. 

The mixing  probability strongly depends on the glueball mass and it turns out to be an analytic function of it,
\begin{align}
c^2 = \frac{1}{2}-\frac{1}{2\sqrt{1+\left(\frac{256}{\Delta'}\right)^2}} \text{ \hspace{3cm}$\eta'$-G system}\\
c^2 = \frac{1}{2}-\frac{1}{2\sqrt{1+\left(\frac{-54}{\Delta}\right)^2}} \text{ \hspace{3cm}$\eta$-G system}
\end{align}
where
\begin{align}
m_G \equiv 960+\Delta' \\
m_G \equiv 550+\Delta.
\end{align}
This comes about from the  diagonalization of the matrices
\begin{align}
\begin{pmatrix}
960 & 128 \\
128 & 960+\Delta'
\end{pmatrix}\text{ \hspace{3cm}$\eta'$-G system}
\\
\begin{pmatrix}
550 & -27 \\
-27 & 550+\Delta
\end{pmatrix}\text{ \hspace{3cm}$\eta$-G system}
\end{align}
when one puts in the angle $\Theta=\arctan{\frac{2\delta}{\Delta}}$, $\delta$ being the mixing energy, into the expression for the mixing probability $c^2=\sin(\frac{\Theta}{2})$.

\begin{figure}[htb]
\begin{center}\label{masses}
\includegraphics[scale=0.9]{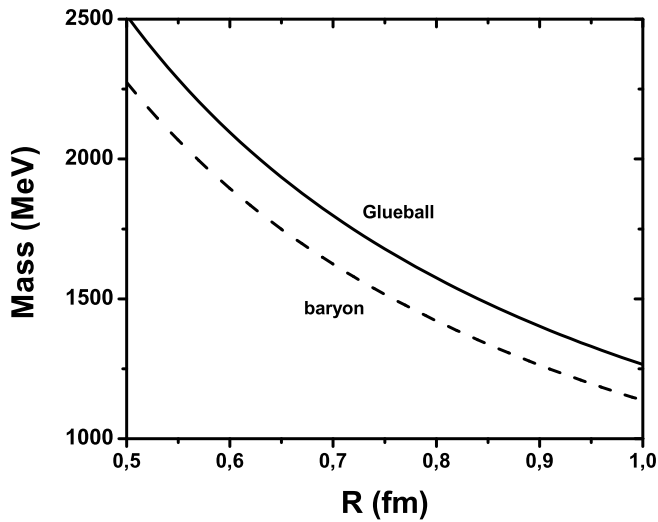}
\end{center}
\caption{Variation of glueball and baryon masses with bag radius.}
\end{figure}

In order to confront the experimental situation let us vary the mass of the glueball following the bag model prescription, i.e.
\begin{equation}
E= \frac{4 (\omega_E + \omega_M - Z)}{3 R},
\label{bagmass}
\end{equation}
where we have eliminated $B$ by the pressure balance equation and we used the the lowest TE and TM modes to calculate the glueball energy. The term Z represents the zero point energy, which we fit to have a glueball mass of $2500$ MeV  at a radius of $0.5$ fm \cite{Kuti:1998rh}. We omit here the perturbative contributions to the mass. This energy has been corrected for center of mass spurious motion to obtain the particle mass which we show in fig. \ref{masses}. Note that the calculation connects the Kuti and Jaffe and Johnson value ranges for different values of the bag radius. We show in the figure the mass of a light baryon calculated with the same zero point energy as a function of radius and see that the it reaches $1100$ MeV at $R = 1.0$ fm, which is the right value before perturbative OGE corrections for the Nucleon-Delta system. Thus we have found a consistent approximate formula to zeroth order which ascribes the value of the glueball mass to its size and which contains all of the results obtained by the different calulcations mentioned above.

\begin{figure}[htb]
\begin{center}
\includegraphics[scale=0.7]{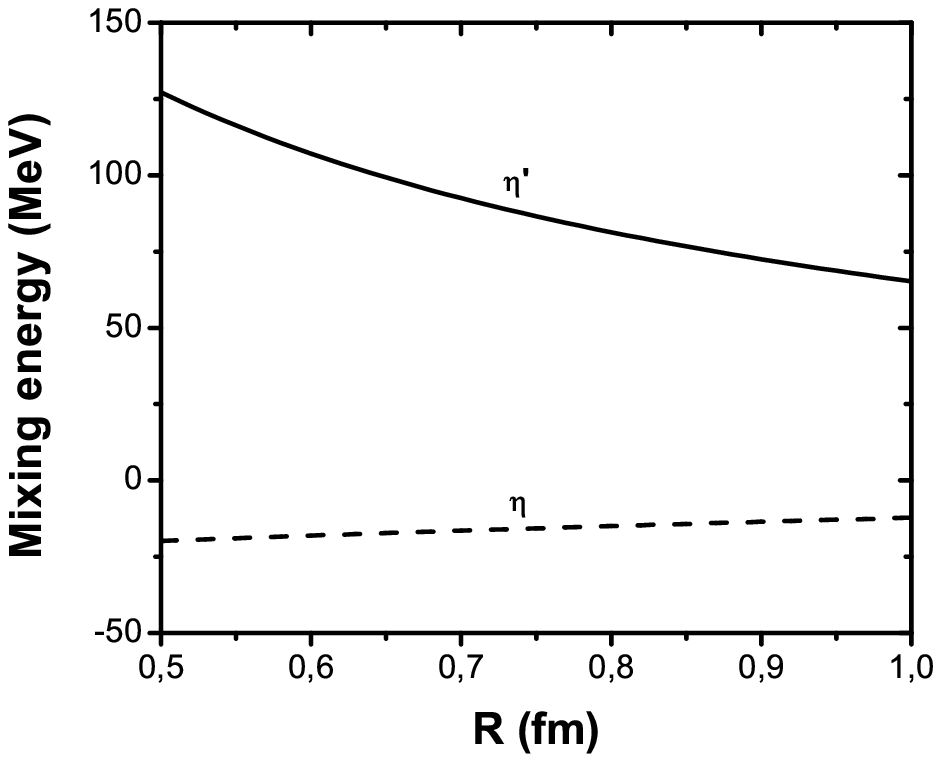}
\includegraphics[scale=0.7]{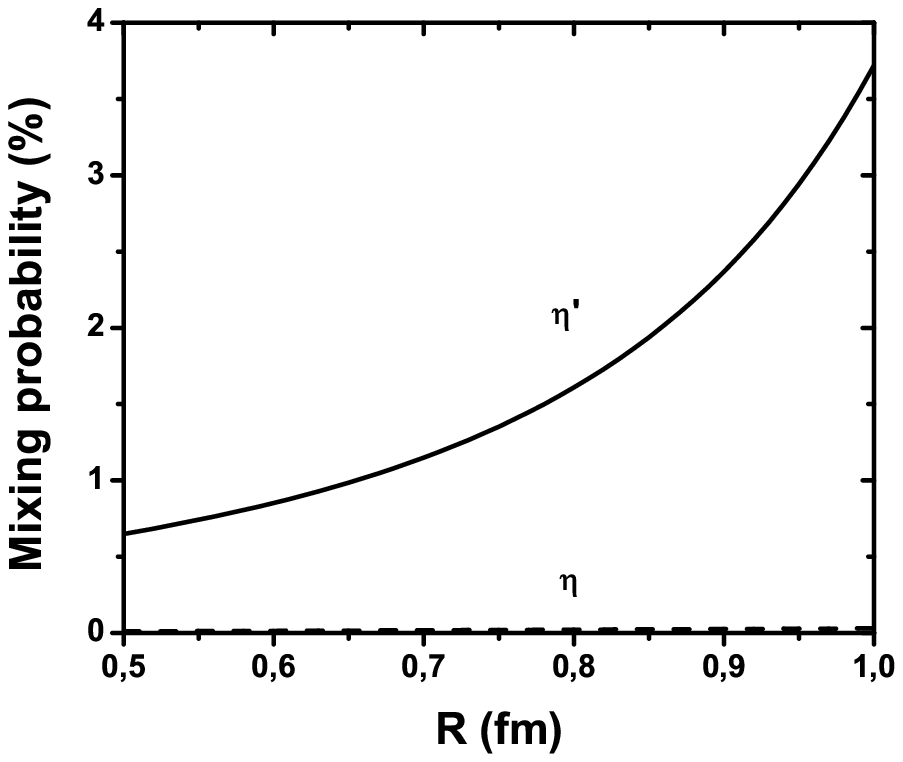}
\caption{$\eta$-glueball and $\eta'$-glueball  mixing energies as a function of bag radius (left)
and $\eta$-glueball and $\eta'$-glueball  probabilities as a function of bag radius (right)}
\end{center}
\label{mixing}
\end{figure}

By using Eq. (\ref{final})  we obtain the mixing energies shown in fig. \ref{mixing}.  From these mixing energies we can calculate  the mixing probability by diagonalizing the mixing Hamiltonian, using for the glueball mass  Eq. \ref{bagmass} and the experimental $\eta$ and $\eta'$ masses. We show them  in fig. \ref{mixing}. We have assumed for the calculation of the mixing energies a constant  $\alpha_S= 0.5$. This is not  the required value to reproduce the baryons and mesons masses at around $ R=1$ fm. The value used in these calculations is closer to $2.0$ \cite{Jaffe:1975fd}. If the glueballs would behave in a similar manner for larger radius, since the mixing energy is proportional to $\alpha_s$, it would increase by a factor of $4$ and the mixing probabilities by a factor of $2-3$. Thus we are showing in the present calculation the minimum values for the mixing probabilities.

It is worth mentioning, that our result deviates significantly from a prior analysis by Carlson and Hansson\cite{Carlson:1981wy}. The difference between our work and theirs is our use the recoil-correction as in \ref{recoil}.  In the figures there is an additional difference arising from the use of different parameters.

The results of fig. \ref{mixing} are quite illuminating. In no case does the $\eta$ mix with the glueball, even for large couplings and small glueball masses. On the contrary the $\eta'$ can mix up to $4\%$ for small glueball masses and up to $12\%$ for large strong $\alpha_S$. We obtain a scenario strongly dependent on the glueball mass. If the glueball mass is close to the lattice value the pseudoscalar glueball should arise as an almost pure state with very distinct features, if on the contrary the mass is small it might mix with the $\eta'$ but never with the $\eta$. 

A caveat we have not considered in our calculation is the mixing between the $\eta_8$ and $\eta_1$ as mentioned before. This mixing would increase the mixing probability of $\eta$-glueball due to its $\eta_1$-component and decrease the mixing of $\eta'$-glueball, due to its $\eta_8$-component. Thus our calculation point  towards a scenario with small mixing of the pseudoscalar glueball with the pseudoscalar mesons.

\section{Conclusions}
\label{conclusion}

We have performed a calculation of the mixing of the pseudoscalar glueball
with the pseudoscalar mesons $\eta$ and $\eta'$. Our work suggests that the
mixing is small if the mass is around 2500 MeV. In the framework of the bag model,
this is a new result. A previous study by Carlson and Hansson\cite{Carlson:1981wy}
suggests a much larger mixing. However, a small glueball mixing, in this glueball mass range,
is as well favored by recent studies \cite{Escribano:2008rq,Cheng:2008ss}.
Accordingly, glueball mixing should not play an important role in the $\eta$-$\eta'$ mass
splitting. This also implies that a rather undiluted, well-defined,
narrow and therefore long-living pseudoscalar glueball state should exist.
If the mass of the glueball is closer to that of the $\eta'$, around $1400$ MeV, 
then the mixing is larger and the consequences of phenomenological analyses 
should be reanalized, since the amount of mixing would determine if the 
glueball behaves more like mesons and baryons, i.e. large objects.

The conceptual difference between our work and other bag model
calculations, like the one carried out by Carlson and Hansson \cite{Carlson:1981wy}, is the
recoil correction. The recoil correction is in our opinion a necessary
ingredient to carry out the calculations in a physically meaningful way
by avoiding spurious singularities. The bag model in the static
spherical cavity approximation fails in describing the creation and
absorption of virtual particles, because in the spherical cavity, the
mode energies are discrete and fixed. Thus energy conservation at the
vertices is generally not possible. This subtle problem has been
neglected in all previous bag model calculations so far. In our
calculation it manifested in an (unphysical) singularity dominating the
results. Our way to resolve this problems is to take into account the recoil that
arises on particle emission and absorption, manifesting itself
necessarily in center-of-mass motion of the constituents. We add this
energy to the energy of the constituents. This not only eliminates the
singularity and makes way for meaningful results, but it is necessary to
cancel out the time-dependence of the mixing energy. Without the recoil
correction, the mixing energy is oscillating, which is also a sign of
center-of-mass motion problems. All of these hints suggest that the recoil
correction is correct, physical meaningful and must be applied to all
bag model calculations.

At this point it is worth mentioning a technical error which despite its
simplicity has been in the field for many years. The normalization constant
eq.\eqref{norm} in the case of massive quarks for the quark mode wavefunction
was written incorrectly,  most probably a typo, in the original paper
~\cite{DeGrand:1975cf}, but the error has been carried on by all papers that
we have used thereafter. Luckily the error does not imply large effects in the
calculations of the light quarks.

In our work, we have neglected the mixing between $\eta$ and $\eta'$ and
rather identified the physical states with the singlet and octet states.
A more elaborated calculation would be a three-particle mixing scheme
between $\eta$, $\eta'$ and the glueball. Nevertheless, since the mixing
between the mesons and the glueball that  we have obtained is so small, a
three-particle mixing scheme will hardly produce very different results
and therefore we have decided to settle for the two-particle mixing scheme.

This work has set the standards for future calculations within the MIT
bag model scheme.  Any calculation dealing with the spectrum or mixings
will have to follow the same procedure and approximations. In particular
an interesting phenomena which we are now revisiting is the mixing of
$\eta_8$ and $\eta_1$ to build the true physical $\eta$-mesons.

\section*{Acknowledgements}
We are grateful to V. Mathieu for interesting discussions. SK thanks the Departamento de F\'{\i}sica Te\'orica of Valencia for the hospitality  and Prof. Kunz-Drolshagen for giving him the opportunity to work there. This work was supported in part  by HadronPhysics2,  a FP7-Integrating Activities and Infrastructure Program of the European Commission
under Grant 227431, by the MICINN (Spain) grant FPA2007-65748-C02-1 and by GVPrometeo2009/129. The diagrams have been drawn using Jaxodraw \cite{Binosi:2003yf}.

\appendix
\section*{Appendix: Details of the calculation}

Substituting eq.\eqref{eq-calc-conf-propagator} into eq.\eqref{eq-thbg-formal-matel} yields a color structure of the form
\begin{align}
\overline{\psi}_i A^a \frac{\lambda^a}{2} \psi_j \overline{\psi}_j A^b \frac{\lambda^b}{2} \psi^C_k
\end{align}
where $\psi^C$ means the charge-conjugated $\psi$.
The gluon color singlet gives a factor of $\delta_{ab}/\sqrt{8}$ and the meson color singlet a factor of $\delta_{ii'}/\sqrt{3}$. This leads to the expression
\begin{align}
\frac{Tr\left(\lambda^a\lambda^a\right)}{4\cdot\sqrt{3}\cdot\sqrt{8}}
\end{align}
Summation over $a$ yields an additional factor of $8$, which finally gives a factor of $\sqrt{\frac{2}{3}}$.

Disregarding the time dependence, we show how to calculate the spatial integrals
\begin{align}
\sum_\alpha \int d^3x \overline{\psi}(x) \slashed{A} u_\alpha(x) \int d^3x' \overline{u}_\alpha(x') \slashed{A}(x')\psi(x')\\
\sum_\alpha \int d^3x \overline{\psi}(x) \slashed{A} v_\alpha(x) \int d^3x' \overline{v}_\alpha(x') \slashed{A}(x')\psi(x')
\end{align}
with the combinations
\begin{align}
\begin{matrix} \overline{\psi}=\overline{u}_\uparrow, \psi=v_\uparrow \\ -\left(\overline{\psi}=\overline{u}_\downarrow, \psi=v_\downarrow\right) \end{matrix}\hspace{1cm} \otimes \hspace{1cm} \begin{matrix} A_1=A_{TE\uparrow},A_2=A_{TM\downarrow} \\ -\left(A_1=A_{TE\downarrow},A_2=A_{TM\uparrow}\right) \\ A_1=A_{TM\downarrow},A_2=A_{TE\uparrow} \\ -\left(A_1=A_{TM\uparrow},A_2=A_{TE\downarrow}\right) \end{matrix}\label{eq-combinations}
\end{align}
We start with the combination $\overline{\psi}=\overline{u}_\uparrow, \psi=v_\uparrow, A=A_{TE\uparrow},A=A_{TM\downarrow}$:
\begin{align}
\sum_\alpha \underbrace{\int d^3x \overline{u}_\uparrow(x) \slashed{A}_{TE\uparrow}(x) u_\alpha(x)}_{(1)}\underbrace{\int d^3 x' \overline{u}_\alpha(x')\slashed{A}_{TM\downarrow}(x')v_\uparrow(x')}_{(2)} \label{eq-conf-matel-sum}
\end{align}
\begin{align}
(1)={}&\int d^3x N_0\left[\begin{pmatrix}-i j_0\left(p_0r\right) \\ \Omega_0 j_{1}\left(p_0r\right)\left(\vec{\sigma}\cdot\hat{\vec{x}}\right)\end{pmatrix}\mathscr{Y}_{0\nicefrac{1}{2}}^{\nicefrac{1}{2}}(\hat{x})\right]^\dagger \gamma_0 \vec{\gamma}\cdot\frac{N_{TE}}{i\omega_{TE}}j_1(\omega_{TE}r)\vec{Y}_{11}^{1}(\hat{x})\nonumber\\
&N_\alpha \begin{pmatrix}i\lambda j_l\left(p_\alpha r\right) \\ \Omega_\alpha j_{l'}\left(p_\alpha r\right)\left(\vec{\sigma}\cdot\hat{\vec{x}}\right)\end{pmatrix}\mathscr{Y}_{lJ}^{M}(\hat{x})
\end{align}
Note that
\begin{align}
\vec{\sigma}\cdot\hat{\vec{x}}\mathscr{Y}_{lJ}^M(\hat{x})=-\mathscr{Y}_{l'J}^M
\end{align}
with $l'=l-\lambda=J-\frac{1}{2}\lambda$ and furthermore
\begin{align}
\gamma_0\vec{\gamma}=\begin{pmatrix}1 & 0 \\ 0 & -1 \end{pmatrix}\begin{pmatrix} 0 & \vec{\sigma} \\ -\vec{\sigma} & 0 \end{pmatrix} = \begin{pmatrix} 0 & \vec{\sigma} \\ \vec{\sigma} & 0 \end{pmatrix}
\end{align}
which yields
\begin{align}
(1)={}&N_0N_\alpha\frac{N_{TE}}{i\omega_{TE}}\int d^3x \begin{pmatrix} i j_0(p_0r)\mathscr{Y}_{0\nicefrac{1}{2}}^{\nicefrac{1}{2}\dagger},-\Omega_0j_1(p_0r)\mathscr{Y}_{1\nicefrac{1}{2}}^{\nicefrac{1}{2}\dagger}\end{pmatrix}\nonumber\\
&\begin{pmatrix}0& \vec{\sigma}\cdot\vec{Y}_{11}^1\\\vec{\sigma}\cdot\vec{Y}_{11}^1 & 0 \end{pmatrix}\cdot\begin{pmatrix}i\lambda j_l\left(p_\alpha r\right) \mathscr{Y}_{lJ}^{M} \\ -\Omega_\alpha j_{l'}\left(p_\alpha r\right)\mathscr{Y}_{l'J}^{M}\end{pmatrix}j_1(\omega_{TE}r)\Omega_\alpha j_{l'}(p_\alpha r)
\end{align}
\begin{align}
={}&N_0N_\alpha\frac{N_{TE}}{i\omega_{TE}}\int d^3x\left[-ij_0(p_0r)\Omega_\alpha j_{l'}(p_\alpha r)\mathscr{Y}_{0\nicefrac{1}{2}}^{\nicefrac{1}{2}\dagger}\vec{\sigma}\cdot\vec{Y}_{11}^1\mathscr{Y}_{l'J}^{M}\right.\nonumber\\
&\left.-\Omega_0j_1(p_0r)i\lambda j_l(p_\alpha r)\mathscr{Y}_{1\nicefrac{1}{2}}^{\nicefrac{1}{2}\dagger}\vec{\sigma}\cdot\vec{Y}_{11}^1\mathscr{Y}_{lJ}^{M}\right]j_1(\omega_{TE}r)
\end{align}
Consider the relation\cite{Varshalovich:1988}
\begin{align}
\mathscr{Y}_{L_1J_1}^{M_1\dagger}\vec{\sigma}\mathscr{Y}_{L_2J_2}^{M_2}={}&(-1)^{J_2+L_1+M_1}\sqrt{\frac{3(2J_1+1)(2J_2+1)(2L_1+1)(2L_2+1)}{2\pi}}\nonumber\\
&\sum_{JL}(-1)^J\left<\begin{array}{cccc|cc}L_10L_20|L0\end{array}\right>\begin{Bmatrix}L_1 & J_1 & \frac{1}{2}\\L_2 & J_2 & \frac{1}{2} \\ L & J & 1 \end{Bmatrix}  \nonumber\\
&\left<\begin{array}{cccc|cc}J_1-M_1J_2M_2|JM\end{array}\right>\vec{Y}_{JM}^L\label{eq-calc-conf-spin-vec}
\end{align}
and furthermore the orthogonality
\begin{align}
\int \vec{Y}_{L'J'}^{M'*}(\hat{x})\cdot\vec{Y}_{LJ}^{M}(\hat{x})d\Omega=\delta_{J'J}\delta_{L'L}\delta_{M'M}
\end{align}
and also
\begin{align}
\vec{Y}_{LJ}^{M*}(\hat{x})=(-1)^{J+L+M+1}\vec{Y}_{LJ}^{-M}(\hat{x}).
\end{align}
These relations allow us to express
\begin{align}
\int \mathscr{Y}_{0\nicefrac{1}{2}}^{\nicefrac{1}{2}\dagger}\vec{\sigma}\cdot\vec{Y}_{11}^1\mathscr{Y}_{l'J}^{M}d\Omega = \int \left[\mathscr{Y}_{0\nicefrac{1}{2}}^{\nicefrac{1}{2}\dagger}\vec{\sigma}\mathscr{Y}_{l'J}^{M}\right]\cdot\vec{Y}_{11}^{-1*}d\Omega
\end{align}
It is now obvious that only the term $J=1,\ L=1$ of the sum in eq.\eqref{eq-calc-conf-spin-vec} survives. Furthermore, $\left<\begin{array}{cccc|cc}L_10L_20|L0\end{array}\right>=\left<\begin{array}{cccc|cc}00l'0|10\end{array}\right>$ tells us that $l'=1$, which allows only  the modes $J=\frac{1}{2},\ \lambda=-1$ and $J=\frac{3}{2},\ \lambda=1$. The factor $\left<\begin{array}{cccc|cc}J_1-M_1J_2M_2|JM\end{array}\right>=\left<\begin{array}{cccc|cc}\frac{1}{2}-\frac{1}{2}JM|1-1\end{array}\right>$ requires $M=-\frac{1}{2}$. \\
The expression
\begin{align}
\int \left[\mathscr{Y}_{1\nicefrac{1}{2}}^{\nicefrac{1}{2}\dagger}\vec{\sigma}\mathscr{Y}_{lJ}^M\right]\cdot Y_{11}^{-1*}d\Omega
\end{align}
requires $l={0,2}$, which allows the modes $J=\frac{1}{2},\ \lambda=-1$, $J=\frac{3}{2},\ \lambda=1$ and $J=\frac{5}{2},\ \lambda=-1$. The last mode, however, is prohibited by the second Clebsch-Gordan coefficient in eq.\eqref{eq-calc-conf-spin-vec}.\\
Thus, the $d\Omega_x$-integral constraints the values for $J,\ \lambda,\ M$ in eq.\eqref{eq-conf-matel-sum}, while the sum goes over all possible $n$.\\
$(1)$ of eq.\eqref{eq-conf-matel-sum} takes the form
\begin{align}
&J=\nicefrac{1}{2},\ \lambda=-1:\nonumber\\
&j_1(\omega_{TE}r)\left[-ij_0(\omega_0r)\Omega_\alpha j_1(\omega_\alpha r)\alpha(0,\nicefrac{1}{2},\nicefrac{1}{2}\parallel 1,\nicefrac{1}{2},-\nicefrac{1}{2}\parallel 1,1)\right.\nonumber\\
&\left.+\Omega_0j_1(\omega_0r)j_0(\omega_\alpha r)\alpha(1,\nicefrac{1}{2},\nicefrac{1}{2}\parallel 0,\nicefrac{1}{2},-\nicefrac{1}{2}\parallel 1,1)\right]
\end{align}
\begin{align}
&J=\nicefrac{3}{2},\ \lambda=1:\nonumber\\
&j_1(\omega_{TE}r)\left[-ij_0(\omega_0r)j_1(\omega_\alpha r)\Omega_\alpha\alpha(0,\nicefrac{1}{2},\nicefrac{1}{2}\parallel 1, \nicefrac{3}{2}, -\nicefrac{1}{2} \parallel 1,1) \right. \nonumber\\
&\left.-i \Omega_0j_1(\omega_0r)j_2(\omega_\alpha r)\alpha(1,\nicefrac{1}{2},\nicefrac{1}{2}\parallel 2,\nicefrac{3}{2},-\nicefrac{1}{2}\parallel 1,1)\right]
\end{align}
where we  left out the expression $N_0N_\alpha\frac{N_{TE}}{i\omega_{TE}}\int dr r^2$ for convenience.\\
We introduce
\begin{align}
\alpha(L_1,J_1,M_1\parallel L_2,J_2,M_2 \parallel L,J) \equiv (-1)^{J+J_2+L_1+M_1}\nonumber\\
\sqrt{\frac{3(2J_1+1)(2J_2+1)(2L_1+1)(2L_2+1)}{2\pi}}\left<\begin{array}{cccc|cc}L_10L_20|L0\end{array}\right>\nonumber\\
\begin{Bmatrix}L_1 & J_1 & \frac{1}{2}\\L_2 & J_2 & \frac{1}{2} \\ L & J & 1 \end{Bmatrix}\left<\begin{array}{cccc|cc}J_1-M_1J_2M_2|JM\end{array}\right>\label{eq-alpha-def}
\end{align}
A list of expressions for $\alpha$ can be found in Table \ref{tab-alpha}\\
\newcommand\T{\rule{0pt}{3.0ex}}
\newcommand\B{\rule[-1.2ex]{1pt}{0pt}}
$(2)$ of eq.\eqref{eq-conf-matel-sum} becomes:
\begin{align}
&J=\nicefrac{1}{2},\ \lambda=-1:\nonumber\\
&\sqrt{\frac{2}{3}}j_0(\omega_{TM}r')\left[ij_0(\omega_\alpha r')j_0(\omega_0r')\alpha(0,\nicefrac{1}{2},-\nicefrac{1}{2}\parallel 0, \nicefrac{1}{2}, \nicefrac{1}{2} \parallel 0, 1) \right. \nonumber\\
&\left.-\Omega_\alpha j_1(\omega_\alpha r')i\Omega_0j_1(\omega r')\alpha(1,\nicefrac{1}{2},-\nicefrac{1}{2}\parallel 1, \nicefrac{1}{2}, \nicefrac{1}{2} \parallel 0,1)\right]\nonumber\\
&-\sqrt{\frac{1}{3}}j_2(\omega_{TM}r')\left[-\Omega_\alpha j_1(\omega_\alpha r')i\Omega_0 j_1(\omega_0 r')\right.\nonumber\\
&\left.\alpha(1, \nicefrac{1}{2}, -\nicefrac{1}{2} \parallel 1, \nicefrac{1}{2}, \nicefrac{1}{2}\parallel 2, 1)\right]\\
\end{align}
\begin{align}
&J=\nicefrac{3}{2},\ \lambda=1\nonumber\\
&\sqrt{\frac{2}{3}}j_0(\omega_{TM}r')\left[-\Omega_\alpha j_1(\omega_\alpha r')i\Omega_0j_1(\omega_0r')\right.\nonumber\\
&\left.\alpha(1,\nicefrac{3}{2},-\nicefrac{1}{2}\parallel 1, \nicefrac{1}{2}, \nicefrac{1}{2} \parallel 0, 1)\right]\nonumber\\
&-\sqrt{\frac{1}{3}}j_2(\omega_{TM}r')\left[-ij_2(\omega_\alpha r')j_0(\omega_0 r')\alpha(2,\nicefrac{3}{2},-\nicefrac{1}{2}\parallel 0, \nicefrac{1}{2}, \nicefrac{1}{2} \parallel 2, 1)\right.\nonumber\\
&\left.-\Omega_\alpha j_1(\omega_\alpha r')i \Omega_0 j_1(\omega_0 r')\alpha(1, \nicefrac{3}{2}, -\nicefrac{1}{2} \parallel 1, \nicefrac{1}{2}, \nicefrac{1}{2} \parallel 2, 1 )\right]
\end{align}
where we left out the expression $-N_0N_\alpha\frac{N_{TM}}{\omega_{TM}}\int dr' r'^2$. The additional minus sign comes from the fact that $\vec{Y}_{01}^{M*}=-\vec{Y}_{01}^{-M}$ and $\vec{Y}_{21}^{M*}=-\vec{Y}_{21}^{-M}$, while $\vec{Y}_{11}^{M*}=\vec{Y}_{11}^{-M}$.

In order for the combinations in \eqref{eq-combinations} to give a nonzero contribution, the spins have to be aligned as $\bar{u}_\uparrow A_\uparrow A_\downarrow v_\uparrow$ or $\bar{u}_\downarrow A_\downarrow A_\uparrow v_\downarrow$. This corresponds to $M_1\rightarrow -M_1, M_2\rightarrow -M_2$ in $\alpha$ \eqref{eq-alpha-def}. Because of
\begin{align}
\frac{\alpha(L_1,J_1,M_1\parallel L_2,J_2,M_2 \parallel L,J)}{\alpha(L_1,J_1,-M_1\parallel L_2,J_2,-M_2 \parallel L,J)}=\left(-1\right)^{J_1+J_2-J+1}\nonumber\\ \text{where }M_1,M_2\in\left\{-\nicefrac{1}{2},\nicefrac{1}{2}\right\}
\end{align}
the signs from the two integrals cancel. Thus, the combinations with inverse magnetic quantum number give the same contribution.\\
Next, we consider the expression
\begin{align}
&\overline{u}_0\vec{\gamma}\cdot \vec{A}_{TE} u_\alpha \overline{u}_\alpha \vec{\gamma}\cdot \vec{A}_{TM} v_0\\
={}&u_0^\dagger\gamma^0\vec{\gamma}\cdot \vec{A}_{TE} u_\alpha u^\dagger_\alpha \gamma^0 \vec{\gamma}\cdot \vec{A}_{TM} i\gamma^2u_0^*
\end{align}
where we inserted the definition for the charge-conjugated particle state. Since this expression has the form of a c-number, we transpose it by Hermitian conjugation and complex conjugation. Hermitian conjugation gives
\begin{align}
u_0^{*\dagger} (-\gamma^2)(-i)(-\vec{\gamma}\cdot\vec{A}_{TM}^*)\gamma^0u_\alpha u_\alpha^\dagger(-\vec{\gamma}\cdot\vec{A}_{TE}^*)\gamma^0u_0.
\end{align}
Complex conjugation gives
\begin{align}
&u_0^{\dagger} \gamma^2 i (\vec{\gamma}\cdot\vec{A}_{TM}-2\left[\vec{A}_{TM}\right]_y\gamma^2)\gamma^0u_\alpha^*\nonumber\\
&u_\alpha^{*\dagger}(\vec{\gamma}\cdot\vec{A}_{TE}-2\left[\vec{A}_{TE}\right]_y\gamma^2)\gamma^0u_0^*\\
={}&-u_0^{\dagger} i (\vec{\gamma}\cdot\vec{A}_{TM})\gamma^2\gamma^0u_\alpha^* u_\alpha^{*\dagger}\gamma^2\gamma^2(\vec{\gamma}\cdot\vec{A}_{TE}-2\left[\vec{A}_{TE}\right]_y\gamma^2)\gamma^0u_0^*\\
={}&-u_0^{\dagger}\gamma^0 (\vec{\gamma}\cdot\vec{A}_{TM})i\gamma^2 u_\alpha^* u_\alpha^{*\dagger}\gamma^2(\vec{\gamma}\cdot\vec{A}_{TE})\gamma^2\gamma^0u_0^*\\
={}&-\overline{u}_0(\vec{\gamma}\cdot\vec{A}_{TM})v_\alpha u_\alpha^{*\dagger}\gamma^2\gamma^0(\vec{\gamma}\cdot\vec{A}_{TE})\gamma^2u_0^*\\
={}&\overline{u}_0(\vec{\gamma}\cdot\vec{A}_{TM})v_\alpha u_\alpha^{*\dagger}\gamma^2\gamma^0i(\vec{\gamma}\cdot\vec{A}_{TE})\gamma^2u_0^*\\
={}&\overline{u}_0(\vec{\gamma}\cdot\vec{A}_{TM})v_\alpha\overline{v}_\alpha(\vec{\gamma}\cdot\vec{A}_{TE})v_0.
\end{align}
For the last step, we have used $\overline{v}=\left(i\gamma^2u^*\right)^\dagger\gamma^0=iu^{*\dagger}\gamma^2\gamma^0$.\\
Taking into account the definition of the confined propagator \eqref{eq-calc-conf-propagator} as well as the time integration with the recoil correction, one finds that the combinations with $TE\leftrightarrow TM$ give the same contribution. Thus, there is an overall symmetry factor of 4 for the different combinations of wavefunctions.

$\overline{u}_0\slashed{A}_{TE}S_F\slashed{A}_{TM}v_0$, particle propagation:
\begin{align}
&J=\nicefrac{1}{2},\ \lambda=-1:\nonumber\\
&\left\{-i\Omega_\alpha\sqrt{\frac{1}{3\pi}}j_0(\omega_0r)j_1(\omega_\alpha r)j_1(\omega_{TE}r)-i\Omega_0\sqrt{\frac{1}{3\pi}}j_1(\omega_0 r)j_0(\omega_\alpha r) j_1(\omega_{TE}r)\right\}\nonumber\\
&\left[\sqrt{\frac{2}{3}}\left\{-i\sqrt{\frac{1}{2\pi}}j_0(\omega_0r')j_0(\omega_\alpha r')j_0(\omega_{TM}r')\right. -i\frac{\sqrt{2}}{6\sqrt{\pi}}\Omega_\alpha\Omega_0j_1(\omega_0r')j_1(\omega_\alpha r')j_0(\omega_{TM}r')\right\}\nonumber\\
&\left.-\sqrt{\frac{1}{3}}\left\{-i\frac{2}{3\sqrt{\pi}}\Omega_\alpha\Omega_0j_1(\omega_0r')j_1(\omega_\alpha r')j_2(\omega_{TM}r')\right\}\right]
\end{align}
\begin{align}
&J=\nicefrac{3}{2},\ \lambda=1:\nonumber\\
&\left\{-i\Omega_\alpha\sqrt{\frac{1}{24\pi}}j_0(\omega_0 r)j_1(\omega_\alpha r)j_1(\omega_{TE}r)+i\Omega_0\sqrt{\frac{1}{24\pi}}j_1(\omega_0r)j_2(\omega_\alpha r)j_1(\omega_{TE}r)\right\}\nonumber\\
&\left[\sqrt{\frac{2}{3}}\left\{-i\frac{1}{3\sqrt{\pi}}\Omega_\alpha\Omega_0j_1(\omega_0r')j_1(\omega_\alpha r')j_0(\omega_{TM}r')\right\}\right. -\sqrt{\frac{1}{3}}\left\{+i\frac{\sqrt{2}}{4\sqrt{\pi}}j_0(\omega_0r')j_2(\omega_\alpha r')j_2(\omega_{TM}r')\right.\nonumber\\
&\left.\left.+i\frac{\sqrt{2}}{12\sqrt{\pi}}\Omega_\alpha\Omega_0j_1(\omega_0r')j_1(\omega_\alpha r')j_2(\omega_{TM}r')\right\}\right]
\end{align}
$\overline{u}_0\slashed{A}_{TE}S_F\slashed{A}_{TM}v_0$, antiparticle propagation:
\begin{align}
&J=\nicefrac{1}{2},\ \lambda=1:\nonumber\\
&\left\{-i\sqrt{\frac{1}{3\pi}}j_0(\omega_0r)j_1(\omega_\alpha r)j_1(\omega_{TE}r)\right. \left.+i\sqrt{\frac{1}{3\pi}}\Omega_0\Omega_\alpha j_1(\omega_0r)j_0(\omega_\alpha r) j_1(\omega_{TE}r)\right\}\nonumber\\
&\left[\sqrt{\frac{2}{3}}\left\{i\sqrt{\frac{1}{2\pi}}\Omega_\alpha j_0(\omega_0 r')j_0(\omega_\alpha r')j_0(\omega_{TM}r')\right. -i\frac{1}{6}\sqrt{\frac{2}{\pi}}\Omega_0j_1(\omega_0r')j_1(\omega_\alpha r')j_0(\omega_{TM}r')\right\}\nonumber\\
&\left.-\sqrt{\frac{1}{3}}\left\{-i\frac{2}{3\sqrt{\pi}}\Omega_0j_1(\omega_0r')j_1(\omega_\alpha r')j_2(\omega_{TM}r')\right\}\right]
\end{align}
\begin{align}
&J=\nicefrac{3}{2},\ \lambda=-1:\nonumber\\
&\left\{i\sqrt{\frac{1}{24\pi}}j_0(\omega_0r)j_1(\omega_\alpha r)j_1(\omega_{TE}r)+i\sqrt{\frac{1}{24\pi}}\Omega_0\Omega_\alpha j_1(\omega_0r)j_2(\omega_\alpha r)j_1(\omega_{TE}r)\right\}\nonumber\\
&\left[\sqrt{\frac{2}{3}}\left\{i\frac{1}{3\sqrt{\pi}}\Omega_0j_1(\omega_0 r')j_1(\omega_\alpha r')j_0(\omega_{TM}r')\right\}\right. -\sqrt{\frac{1}{3}}\left\{-i\Omega_\alpha \frac{\sqrt{2}}{4\sqrt{\pi}}j_0(\omega_0 r')j_2(\omega_\alpha r')j_2(\omega_{TM}r')\right.\nonumber\\
&\left.\left. -i \Omega_0\frac{\sqrt{2}}{12\sqrt{\pi}}j_1(\omega_0r')j_1(\omega_\alpha r') j_2(\omega_{TM}r')\right\}\right]
\end{align}
For convenience, we have left out the expression $-i\displaystyle\sum_n N_0^2 N_\alpha^2\frac{N_{TE}N_{TM}}{i\omega_{TE}\omega_{TM}}\int dr r^2 \int dr' r'^2$. Also, one has a factor of $2g^2\cdot4\cdot\frac{1}{2}\cdot\sqrt{\frac{2}{3}}$, which arises from eq.\eqref{eq-thbg-formal-matel}, the symmetry in the combinations, the wavefunction symmetrization and the color matrix trace, respectively.
For the confined propagator, it is important to note the $J,\ \lambda$ quantum numbers, because the modes $\omega_\alpha$ depend on these quantum numbers as well as on the $n$ quantum number, which is summed over.
\begin{table}
\centering
\begin{tabular}{ccc|ccc|cc||cc}
$L_1$ & $J_1$ & $M_1$ & $L_2$ & $ J_2$ & $M_2$ & $L$ & $J$ & $\alpha$ \\\hline
$1$ & $\nicefrac{1}{2}$ & $\nicefrac{1}{2}$ & $0$ & $\nicefrac{1}{2}$  & $-\nicefrac{1}{2}$ & $1$ & $1$ &  $ -\sqrt{\frac{1}{3\pi}}$\\[8pt]
$0$ & $\nicefrac{1}{2}$ & $\nicefrac{1}{2}$ & $1$ & $\nicefrac{3}{2}$ & $-\nicefrac{1}{2}$ & $1$ & $1$ & $\sqrt{\frac{1}{24\pi}}$ \\[8pt]
$1$ & $\nicefrac{1}{2}$ & $\nicefrac{1}{2}$ & $2$ & $\nicefrac{3}{2}$ & $-\nicefrac{1}{2}$ & $1$ & $1$ & $-\sqrt{\frac{1}{24\pi}}$\\[8pt]
$0$ & $\nicefrac{1}{2}$ & $-\nicefrac{1}{2}$ & $0$ & $\nicefrac{1}{2}$ & $\nicefrac{1}{2}$ & $0$ & $1$ & $-\sqrt{\frac{1}{2\pi}}$\\[8pt]
$1$ & $\nicefrac{3}{2}$ & $-\nicefrac{1}{2}$ & $1$ & $\nicefrac{1}{2}$ & $\nicefrac{1}{2}$ & $2$ & $1$ & $-\frac{\sqrt{2}}{12\sqrt{\pi}}$\\[8pt]
$1$ & $\nicefrac{1}{2}$ & $-\nicefrac{1}{2}$ & $1$ & $\nicefrac{1}{2}$ & $\nicefrac{1}{2}$ & $2$ & $1$ & $\frac{2}{3\sqrt{\pi}}$\\[8pt]
$1$ & $\nicefrac{3}{2}$ & $-\nicefrac{1}{2}$ & $1$ & $\nicefrac{1}{2}$ & $\nicefrac{1}{2}$ & $0$ & $1$ & $\frac{1}{3\sqrt{\pi}}$\\[8pt]
$0$ & $\nicefrac{1}{2}$ & $\nicefrac{1}{2}$ & $2$ & $\nicefrac{3}{2}$ & $-\nicefrac{1}{2}$ & $2$ & $1$ & $\frac{\sqrt{2}}{4\sqrt{\pi}}$\\[8pt]
$1$ & $\nicefrac{1}{2}$ & -$\nicefrac{1}{2}$ & $1$ & $\nicefrac{1}{2}$ & $\nicefrac{1}{2}$ & $0$ & $1$ & $\frac{\sqrt{2}}{6\sqrt{\pi}}$
\end{tabular}
\caption{$\alpha$ coefficients}
\label{tab-alpha}
\end{table}


\newpage
\begin{thebibliography}{99}

\bibitem{FritzschGellMannLeutwyler}
H. Fritzsch, M. Gell-Mann and H. Leutwyler, {\it Phys. Lett.} B47 (1973) 365.


\bibitem{Gross:1973id}
D.~J.~Gross and F.~Wilczek,
{\it Phys.\ Rev.\ Lett.\ }  {\bf 30} (1973) 1343.

\bibitem{Politzer:1973fx}
H.~D.~Politzer,
{\it Phys.\ Rev.\ Lett.\  }{\bf 30} (1973) 1346.

\bibitem{Wilson:1974sk}
K.~G.~Wilson,
{\it Phys.\ Rev.\  D } {\bf 10} (1974) 2445.



\bibitem{Mathieu:2008me}
V.~Mathieu, N.~Kochelev and V.~Vento,
Int.\ J.\ Mod.\ Phys.\  E {\bf 18} (2009) 1
[arXiv:0810.4453 [hep-ph]].


\bibitem{Crede:2008vw}
V.~Crede and C.~A.~Meyer,
arXiv:0812.0600 [hep-ex].

\bibitem{Scharre:1980zh}
D.~L.~Scharre {\it et al.},
Phys.\ Lett.\  B {\bf 97} (1980) 329.

\bibitem{FritzschMinkowski}
H. Fritzsch and P. Minkowski, {\it Nuov. Cim. }{\bf 30}A (1975) 393.


\bibitem{Jaffe:1975fd}
R.~L.~Jaffe and K.~Johnson,
{\it Phys.\ Lett.\  B }{\bf 60}, 201 (1976).

\bibitem{ozi} S. Okubo, {\it Phys. Lett. }{\bf 5}, 1975 (1963);
G. Zweig, in {\it Development in the Quark Theory of Hadrons}, edited by D.B. Lichtenberg and
S.P. Rosen (Hadronic Press, Massachusetts, 1980); J. Iizuka, {\it Prog. Theor. Phys. Suppl. }
{\bf 37}, 38 (1966).


\bibitem{Maxwell:1981kg}
O.~V.~Maxwell and V.~Vento,
Nucl.\ Phys.\  A {\bf 407} (1983) 366.


\bibitem{Messiah62}
A. Messiah, Mécanique Quantique, vol. 2, (Dunod, Paris 1962).


\bibitem{DeGrand:1975cf}
T.~A.~DeGrand, R.~L.~Jaffe, K.~Johnson and J.~E.~Kiskis,
Phys.\ Rev.\  D {\bf 12} (1975) 2060.

\bibitem{Morningstar:1999rf}
C.~J.~Morningstar and M.~J.~Peardon,
{\it Phys.\ Rev.\  D} {\bf 60}, 034509 (1999)
[arXiv:hep-lat/9901004].

\bibitem{Meyer:2004jc}
H.~B.~Meyer and M.~J.~Teper,
{\it Phys.\ Lett.\  B} {\bf 605} (2005) 344
[arXiv:hep-ph/0409183].
H.~B.~Meyer,
arXiv:hep-lat/0508002.

\bibitem{Chen:2005mg}
Y.~Chen {\it et al.},
{\it Phys.\ Rev.\  D} {\bf 73} (2006) 014516
[arXiv:hep-lat/0510074].

\bibitem{Sexton:1995kd}
J.~Sexton, A.~Vaccarino and D.~Weingarten,
{\it Phys.\ Rev.\ Lett.\ }{\bf 75} (1995) 4563
[arXiv:hep-lat/9510022].

\bibitem{Mathieu:2009sg}
V.~Mathieu and V.~Vento,
arXiv:0910.0212 [hep-ph].

\bibitem{Cheng:2008ss}
H.~Y.~Cheng, H.~n.~Li and K.~F.~Liu,
Phys.\ Rev.\  D {\bf 79} (2009) 014024
[arXiv:0811.2577 [hep-ph]].

\bibitem{Kuti:1998rh}
J.~Kuti,
Nucl.\ Phys.\ Proc.\ Suppl.\  {\bf 73} (1999) 72
[arXiv:hep-lat/9811021].


\bibitem{Carlson:1981wy}
C.~E.~Carlson and T.~H.~Hansson,
Nucl.\ Phys.\  B {\bf 199} (1982) 441.


\bibitem{Escribano:2008rq}
R.~Escribano,
Eur.\ Phys.\ J.\  C {\bf 65} (2010) 467
[arXiv:0807.4201 [hep-ph]].



\bibitem{Binosi:2003yf}
D.~Binosi and L.~Theussl,
Comput.\ Phys.\ Commun.\  {\bf 161} (2004) 76
[arXiv:hep-ph/0309015].


\bibitem{Varshalovich:1988}
D. A. Varshalovich, A.N. Moskalev, V. K. Khersonskii, Quantum Theory of Angular Momentum,
(World Scientific, Singapore 1988).


\end{thebibliography}
\end{document}